\documentclass{nature}

\usepackage{graphicx}
\usepackage{subfigure}
\usepackage{amsmath}
\usepackage{amssymb}
\usepackage{aas_macros}
\usepackage{float}

\title{Collisionless shock heating of heavy ions in SN 1987A}

\author{Marco Miceli$^{*,1,2}$, Salvatore Orlando$^{2}$, David N. Burrows$^3$, Kari A. Frank$^4$, Costanza Argiroffi$^{1,2}$, Fabio Reale$^{1,2}$, Giovanni Peres$^{1,2}$, Oleh Petruk$^5$, Fabrizio Bocchino$^{2}$}

\begin{document}

\maketitle

\begin{affiliations}
 \item Dipartimento di Fisica e Chimica, Universit\`a degli Studi di Palermo, Piazza del Parlamento 1, 90134 Palermo, Italy
 \item INAF-Osservatorio Astronomico di Palermo, Piazza del Parlamento 1, 90134 Palermo, Italy, 
 \item Department of Astronomy and Astrophysics, Pennsylvania State University, University Park, PA 16802, USA
 \item Northwestern University, Technological Institute, 2145 Sheridan Road, Evanston, IL 60208-3112, USA
 \item Institute for Applied Problems in Mechanics and Mathematics, Naukova Str. 3-b, UA-79060 Lviv, Ukraine
\end{affiliations}

\begin{abstract}

Astrophysical shocks at all scales, from those in the heliosphere up to the cosmological shock waves, are typically ``collisionless", because the thickness of their jump region is much shorter than the collisional mean free path. Across these jumps, electrons, protons, and ions are expected to be heated at different temperatures. Supernova remnants (SNRs) are ideal targets to study collisionless processes because of their bright post-shock emission and fast shocks. Although optical observations of Balmer-dominated shocks in young SNRs showed that the post-shock proton temperature is higher than the electron temperature, the actual dependence of the post-shock temperature on the particle mass is still widely debated\cite{gsm13}.
We tackle this longstanding issue through the analysis of deep multi-epoch and high-resolution observations of the youngest nearby supernova remnant, SN 1987A, made with the \emph{Chandra} X-ray telescope. We introduce a novel data analysis method by studying the observed spectra in close comparison with a dedicated full 3-D hydrodynamic simulation. The simulation is able to reproduce self-consistently the whole broadening of the spectral lines of many ions altogether. We can therefore measure the post shock temperature of protons and selected ions through comparison of the model with observations. We have obtained information about the heating processes in collisional shocks by finding that the ion to proton temperature ratio is always significantly higher than one and increases linearly with the ion mass for a wide range of masses and shock parameters.
\end{abstract}

Shock waves are abrupt transitions between a supersonic and a subsonic flow which transform bulk kinetic energy into thermal energy by compressing and heating the medium. In the Earth's atmosphere the width of a shock front is of the order of a few molecular mean free paths. In the rarefied astrophysical environments, however, particle$-$particle interactions (Coulomb collisions) are typically not sufficient to provide the viscous dissipation, and collective effects, such as electromagnetic fluctuations and plasma waves, provide the Rankine-Hugoniot\cite{ll59} jump conditions at the shock front\cite{bdd08,vin12}. These conditions derive from the mass, momentum, and energy conservation across the shock and predict that the post-shock temperature $T$ depends on the shock velocity $v_s$ as $kT=3/16~mv_{s}^{2}$, where $m$ is the particle mass. In a plasma with different particle species, it is still not clear whether a (partial) temperature equilibration between different species can be reached, or particles with different masses reach temperatures proportional to their mass as
\begin{equation}
kT_i=\frac{3}{16} m_i v_{s}^{2}
\label{eq:tshock}
\end{equation}
where $m_i$ is the particle mass for the i-th species. Collisionless shocks have been observed decades ago in the solar wind\cite{ts85}, as well as on cosmological scales\cite{bdd08}.
A post-shock temperature proportional to the particle mass is expected in case of scattering isotropization of the incoming particles by plasma waves. Nevertheless, partial equilibrium between different species is also possible and the validity of equation (\ref{eq:tshock}) is far from being settled. Thus, the actual conditions of the post-shock plasma are still under debate.

Pioneering works have shown the importance of Balmer-dominated shock fronts in SNRs as diagnostic tools\cite{cr78,ckr80} and the study of the H$\alpha$ line profile is widely used to measure the electron to proton temperature ratio $T_e/T_p$\cite{gsm13,ray18}. However, this ratio is typically much higher than the electron to proton mass ratio ($m_e/m_p$) and can increase up to 1 in slow ($v_s\sim400$ km s$^{-1}$) shocks\cite{rgh03,vhm08}, showing a dependence on the shock velocity which has been modelled as $T_{e}/T_{p}\propto v_{s}^{-2}$. This can be explained if the immediate electron post-shock temperature does not depend on the shock velocity\cite{glr07} (and is always $kT_e\sim0.3$ keV), while $T_p$ varies as in equation (\ref{eq:tshock}). This behaviour can be associated with a mechanism of electron heating due to lower hybrid waves in the shock precursor\cite{rlg08}, though other scenarios have been proposed\cite{vbb15}. The general expectation is that there can be different plasma instabilities that can enhance $T_{e}/T_{p}$ above the expected value $m_e/m_p$ and the electron heating processes in collisionless shocks are different from those of ions\cite{sh00,bdd08,pcs15}.  Therefore, it is necessary to accurately measure the ion temperatures to test equation (\ref{eq:tshock}).

However, the measurement of the post shock temperatures for different ions has produced contradictory results\cite{ray18}: the temperature of oxygen ions relative to protons was found to be less than that predicted by equation (\ref{eq:tshock}) in the analysis of UV observations of SN 1006\cite{krz04}, while it was found to be higher than equation (\ref{eq:tshock}) predictions in interplanetary shocks\cite{bgg97}.
Recently, an important result has been obtained in SN 1006, with He, C and N ion temperatures being consistent with the mass-proportional scenario\cite{rwb17}.
Stronger constraints need to be obtained by testing equation (\ref{eq:tshock}) over a wider range of masses, by inspecting elements heavier than N, or O. To this end, the X-ray band is the ideal window in which bright emission lines of heavy ions are typically observed. 
Up to now, only in one case X-ray spectra were used to measure a line broadening of the OVII line triplet, corresponding to an extremely high oxygen temperature ($\sim300$ keV) in an isolated ejecta knot of SN 1006\cite{bvm13,vlg03}.

SN 1987A in the Large Magellanic Cloud offers a unique opportunity of observing a nearby, young, bright, SNR with high level of detail. SN 1987A was a hydrogen-rich core-collapse SN discovered on 1987 February 23\cite{wls87}. 
Its evolution has been extensively covered by a wealth of observations in different wavelengths\cite{mcr93,mcf16} and reveals a complex interaction of the blast wave with the surrounding inhomogeneous medium, characterized by a dense, clumpy ring-like nebula, inside a more diffuse HII region.
The interaction with the nebula is best revealed in X-rays and SN 1987A has been monitored through dedicated campaigns of observations with {\it XMM-Newton} and {\it Chandra}.

The series of X-ray observations encodes information about the physical properties of both the nebula and the stellar ejecta and requires a thorough data analysis: phenomenological models only analyzed single observations, regardless of the whole succession of data sets.
Here, in our novel approach, a single 3-D hydrodynamic model\cite{omp15} describes the evolution of SN 1987A from the onset of the supernova to the current age and accounts self-consistently for all the observations and for the evolution of the system.
The reliability of the model has been tested and confirmed by synthesizing light curves, images, and low-resolution (CCD) spectra from the hydrodynamic simulations. We found that our model self-consistently fits: i) the bolometric light curve during the first 250 days of evolution, ii) the soft ($0.5-2$ keV) and hard ($3-10$ keV) X-ray light curves in the subsequent 30 years, iii) the evolution of the morphology of the X-ray emission, iv) {\it XMM-Newton} EPIC and {\it Chandra} ACIS spectra at different epochs\cite{omp15}. 
Here we use this forward modeling approach to obtain deeper insight into the physics of shock heating, through the detailed reproduction of the multi-epoch high-resolution gratings X-ray observations of SN 1987A.

To synthesize the $Chandra$ gratings spectra from the model, we included all the sources of line broadening, namely: i) the bulk velocities of the different parts of the ring, ii) the instrumental broadening (due to both line response function and X-ray source extension in the dispersion direction), and iii) the thermal broadening. To add the contribution of thermal heating in our synthetic spectra, we assumed the ion temperature to be mass-proportional with respect to the proton temperature. The latter is accurately followed by our hydrodynamic code, which includes a detailed model of shock heating\cite{glr07} and post-shock evolution due to Coulomb collisions between protons and electrons\cite{spi62}. We produced synthetic spectra either with or without thermal broadening. By comparing the synthetic line widths with those measured in the actual spectra, we were able to infer the role of thermal broadening and its dependence on the ion mass.

We considered the two deepest observations of SN~1987A performed with the MEG spectrometer of the {\it Chandra} High Energy Transmission Grating (HETG). The two data sets mark two different evolutionary states and different conditions in the shocked plasma: the first one consists of a set of 14 exposures performed between March and April 2007 (SN days 7321$-$7358, total exposure time of 354.9 ks) and corresponds to the initial phase of interaction between the blast-wave and the circumstellar ring, while in the second one (4 HETG exposures, March 2011, 178 ks) the shock has already crossed the bulk of the ring and the X-ray flux is $\sim2$ times higher than in 2007\cite{fzp16}.

\begin{figure*}[ht!]
\begin{minipage}{0.5\textwidth}
 \includegraphics[width=.99\textwidth]{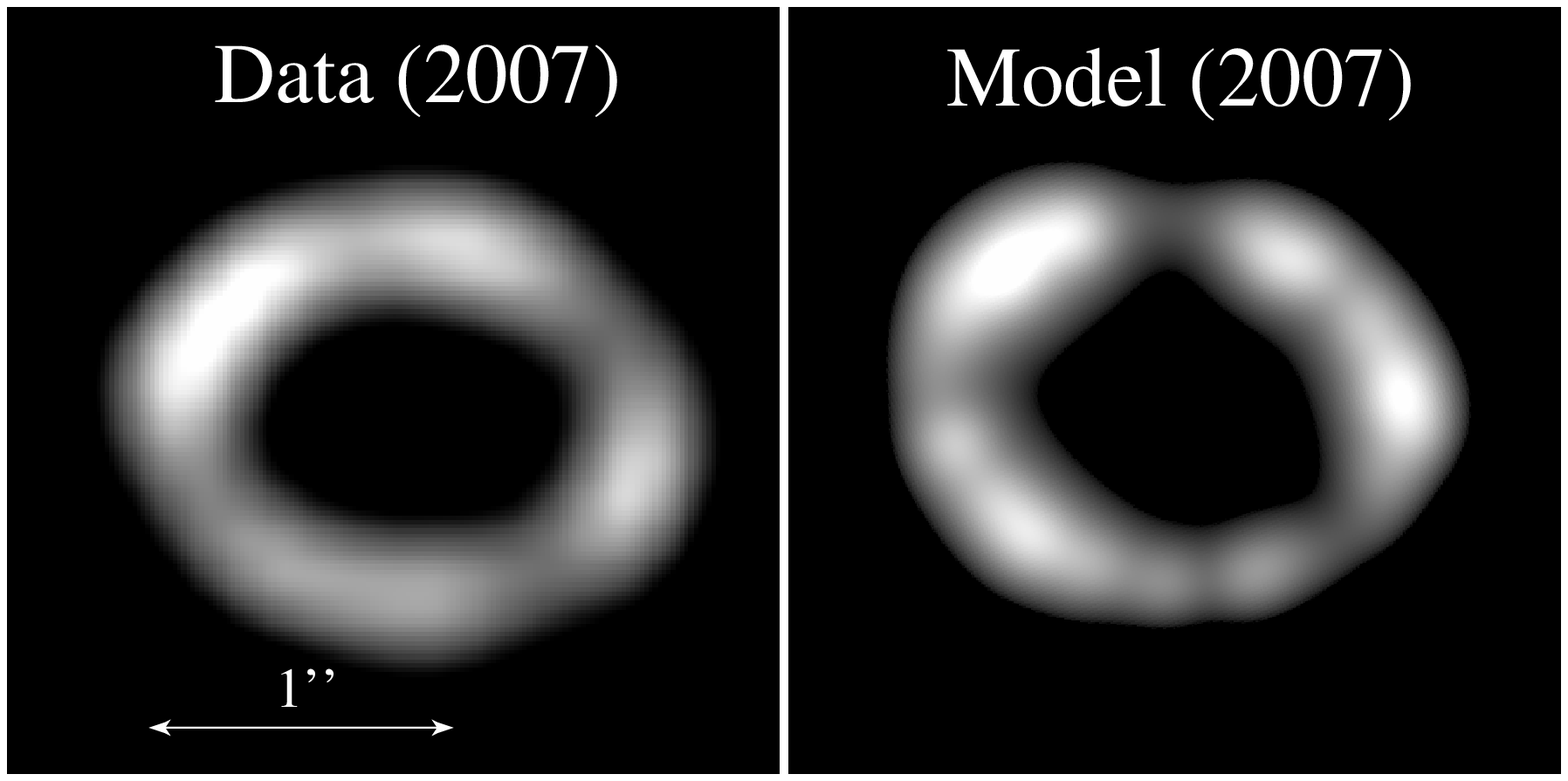}
 \end{minipage}
 \hfill
 \begin{minipage}{0.5\textwidth}
 \includegraphics[width=.99\textwidth]{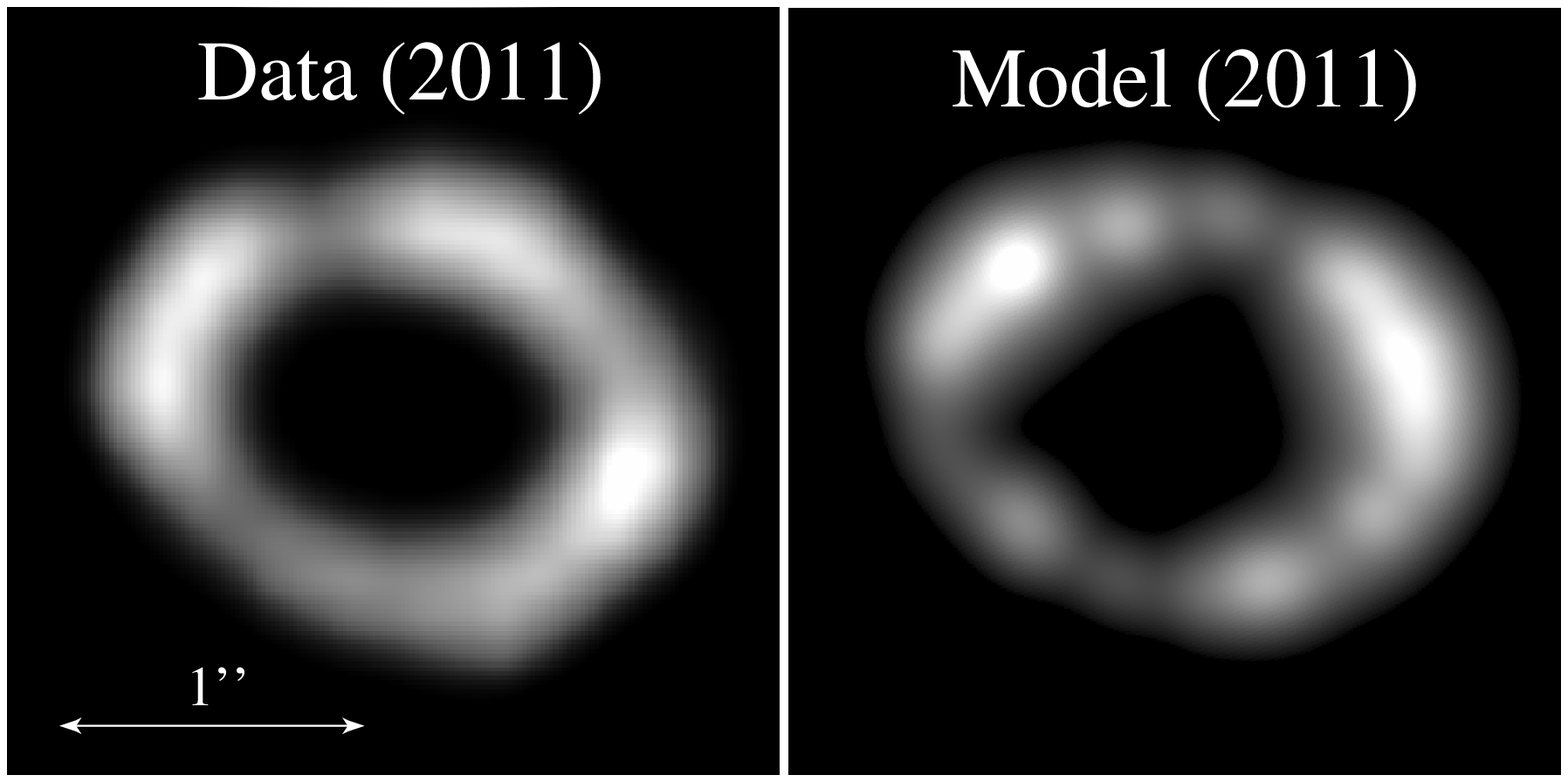}
 \end{minipage}
\begin{minipage}{0.5\textwidth}
 \includegraphics[angle=90,width=.99\textwidth]{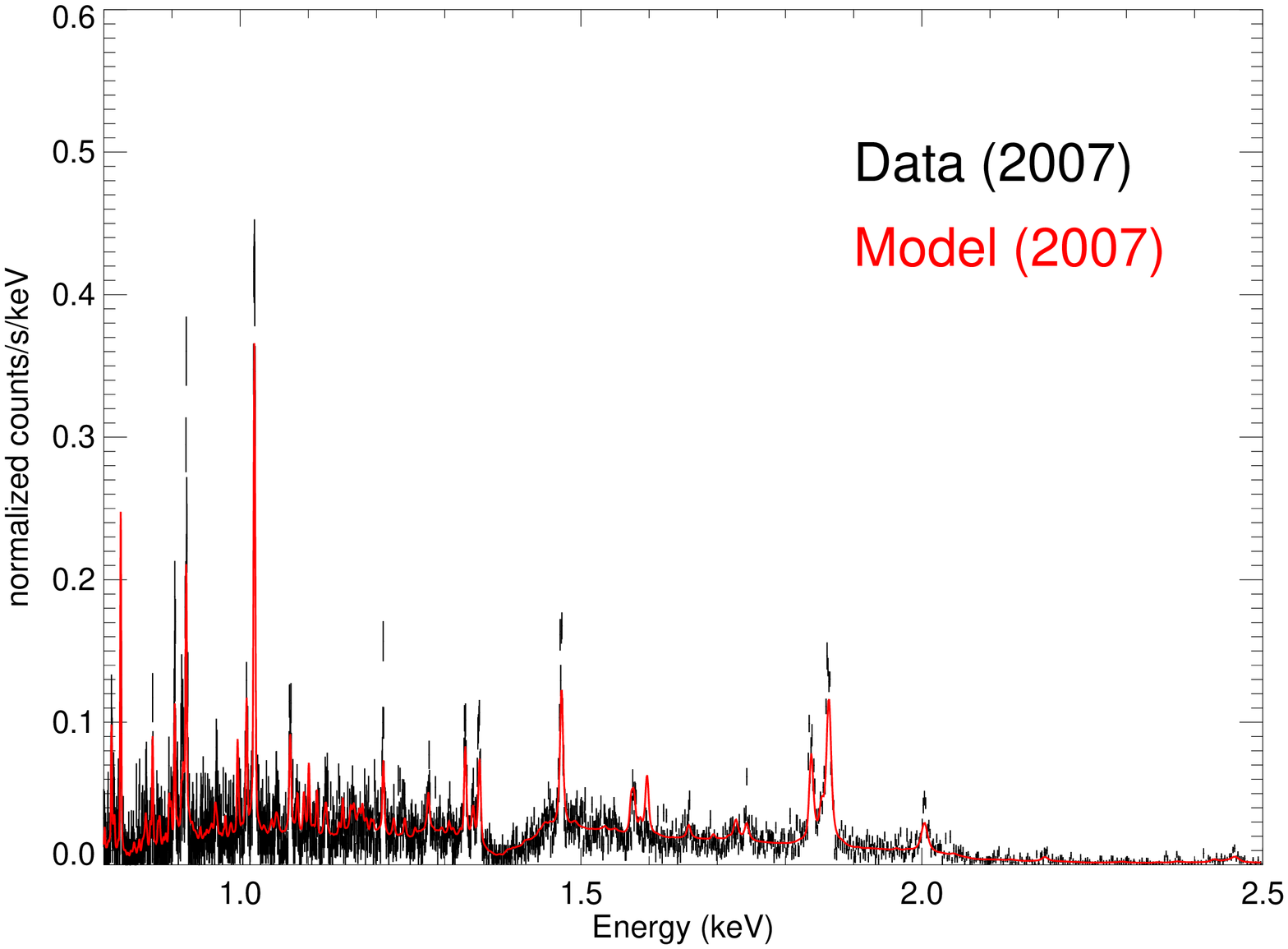}
 \end{minipage}
 \hfill
 \begin{minipage}{0.5\textwidth}
 \includegraphics[angle=90,width=.99\textwidth]{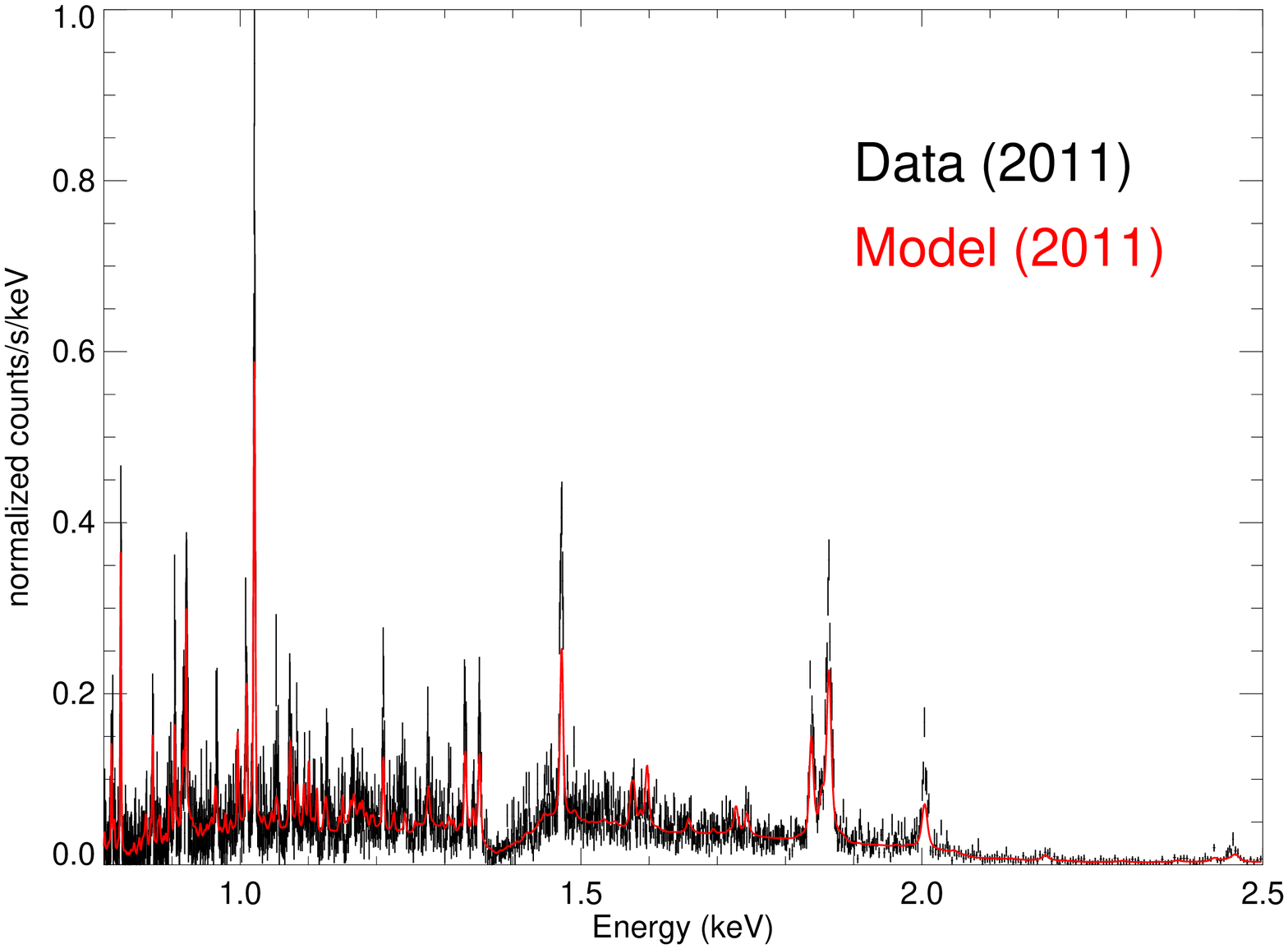}
 \end{minipage}
 \begin{minipage}{0.5\textwidth}
 \includegraphics[angle=90,width=.99\textwidth]{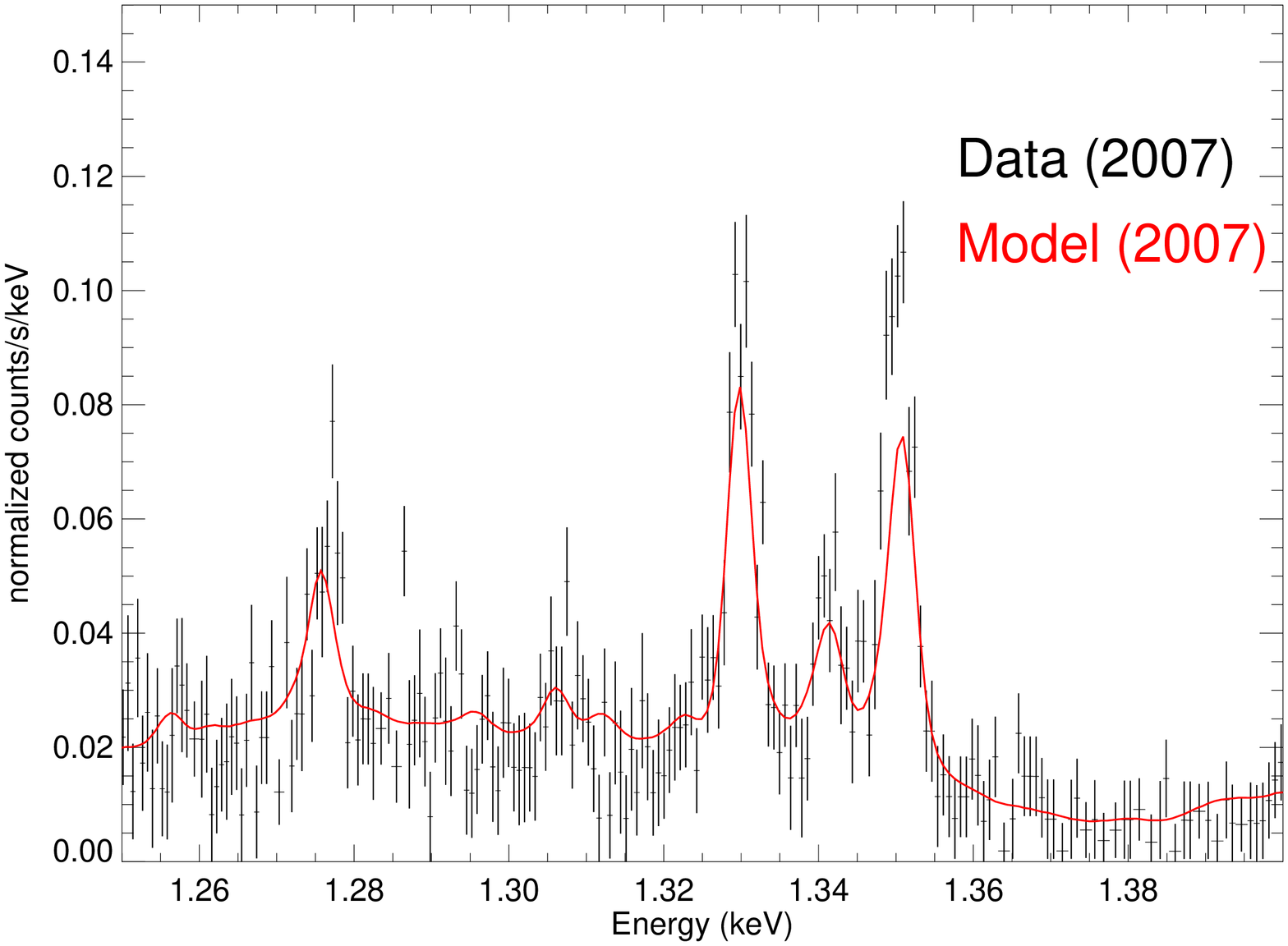}
 \end{minipage}
 \hfill
 \begin{minipage}{0.5\textwidth}
 \includegraphics[angle=90,width=.99\textwidth]{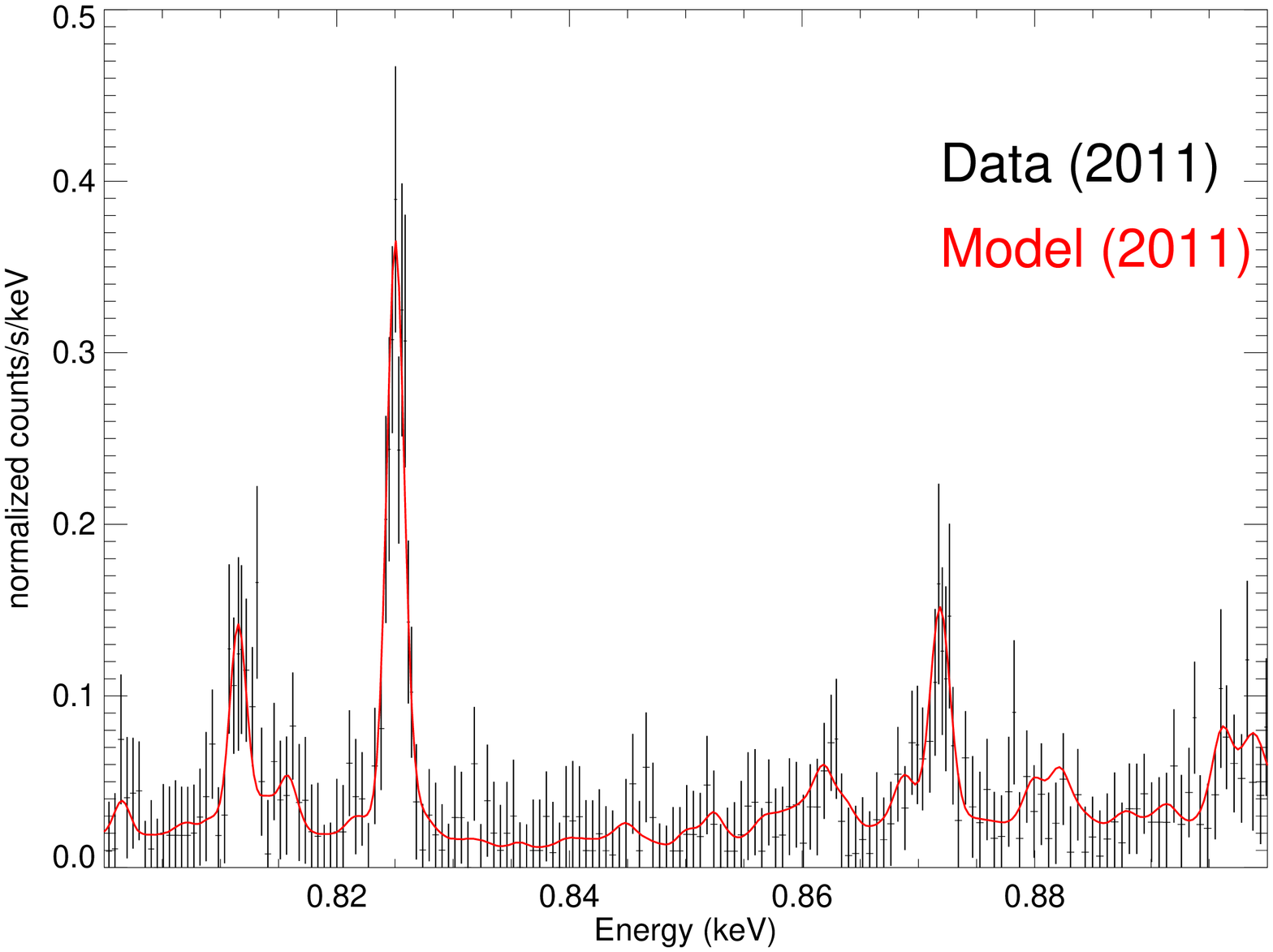}
\end{minipage}
 \caption{\emph{Upper panels:} Observed and synthetic maps of SN 1987A in the $0.5-2$ keV band in 2007 (left) and 2011 (right). \emph{Central panels:} Observed (black) and synthetic (red, with all sources of line broadening) X-ray spectra of SN 1987A in the $0.8-2.5$ keV band (MEG $+1$ order of the {\it Chandra} HETG) in 2007 (left) and 2011 (right). \emph{Lower panels:} close-up views of the central panels in the $1.25-1.4$ keV energy band (in 2007, left) and $0.8-0.9$ keV band (in 2011, right).}
\label{fig:imgspec}
\end{figure*}

The hydrodynamic model simulates the evolution of the ring throughout this time range. Upper panels of Fig. \ref{fig:imgspec} show the observed and synthetic X-ray images of SN 1987A, while central panels show the comparison between the observed high resolution spectra of SN 1987A in 2007 and 2011 and the synthetic spectra derived from our hydrodynamic simulations including all the possible sources of line broadening. The model agrees closely with the observations and matches in detail the X-ray spectra in both epochs, where emission lines from Fe XVII and He-like and H-like ions of Ne, Mg, and Si are visible. 

\begin{figure*}[tb!]
\centering
\includegraphics[angle=90,width=0.49\textwidth]{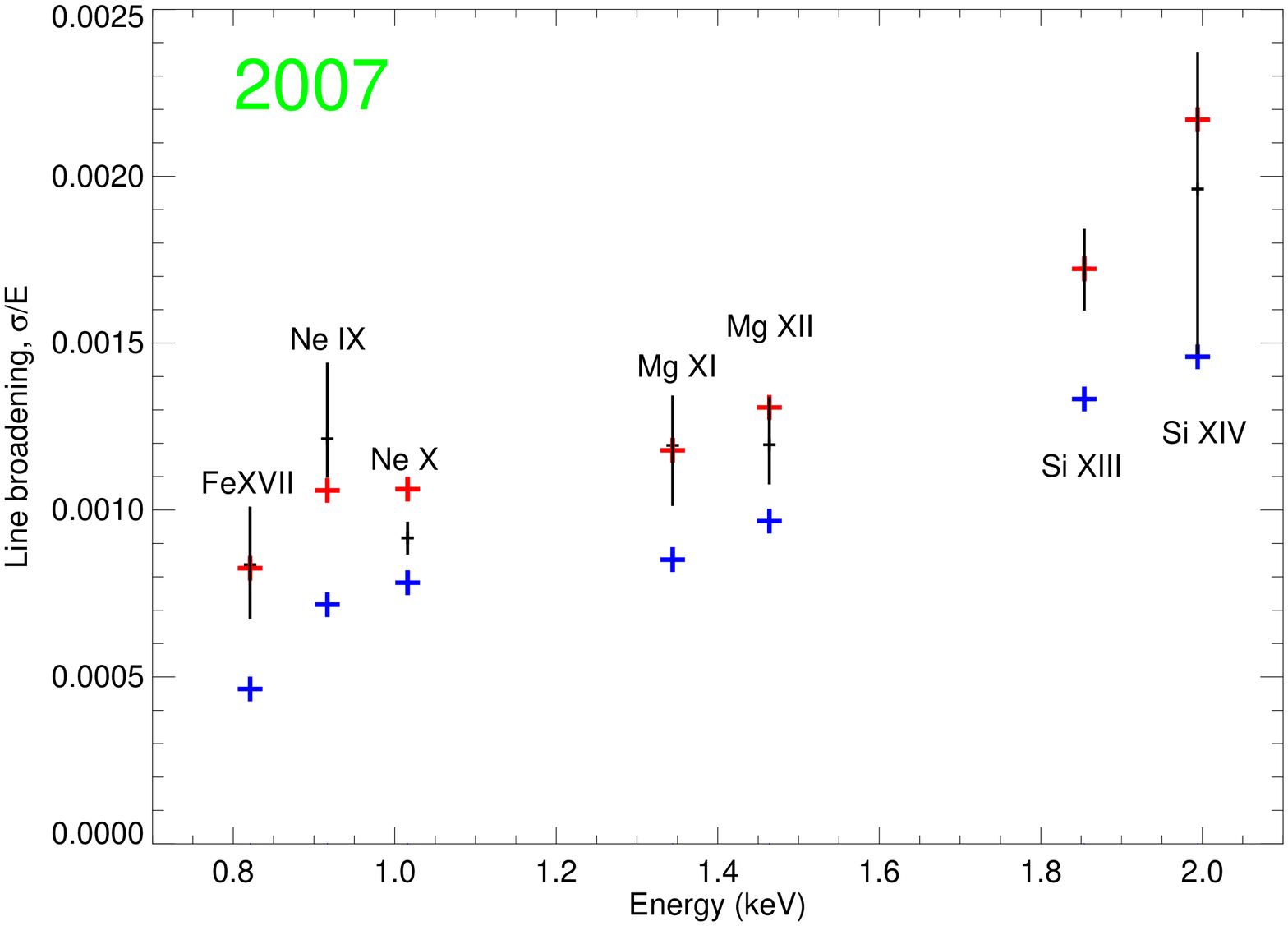}
\includegraphics[angle=90,width=0.49\textwidth]{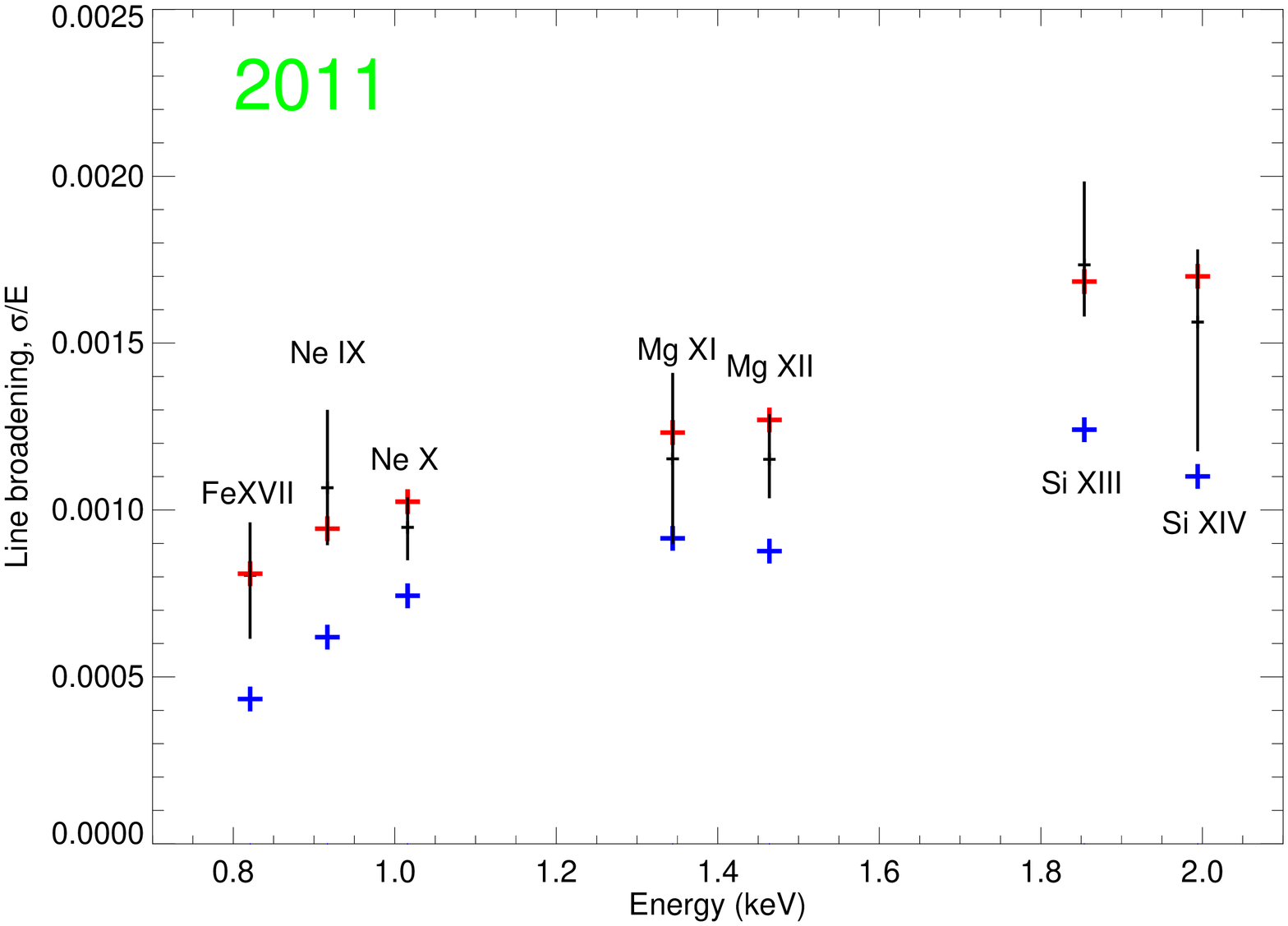}
\caption{Line broadening for selected ions in the 2007 ({\it left panel}) and 2011 ({\it right panel}) X-ray spectra of SN 1987A: black crosses show the observed values (MEG $+1$ order of the {\it Chandra} HETG), while blue and red crosses indicate the values synthesized from our hydrodynamic model by considering only Doppler and instrumental broadening, or Doppler, instrumental, and thermal broadening, respectively. Error bars are at the 90$\%$ confidence level.}
\label{fig:broad}
\end{figure*}

The model reproduces even the significant line broadening of the single lines (lower panels of Fig. \ref{fig:imgspec}), which results from a combination of Doppler effects, due to the bulk velocities of the approaching and receding parts of the ring, thermal broadening, associated with the high temperatures of ions, and instrumental effects. Previous works neglected thermal broadening and used the line widths to derive {\it a posteriori} the bulk velocity of the plasma\cite{zmb05,zmd09,ddh12}. Our model provides us with complete information to derive the total line broadening and its evolution in a self-consistent way, i.~e., all the hydrodynamics and thermodynamics, and the bulk motion of the shocked plasma at all times.
In particular, the Doppler broadening depends on the ring and clump densities (which affect the post-shock dragging) and on the density/velocity profiles of the outer ejecta; all these parameters are very well constrained by our model.
To highlight the contribution of thermal broadening, we compare in Fig. \ref{fig:broad} the line broadening derived from the model (with and without thermal contribution) with the observed line widths for the two data sets.
The figure clearly shows that the line broadening from the model without thermal contribution is large and changes from line to line and in time, as the shock expands through different parts of the ring. Nevertheless, it is always significantly smaller than the observed one, i.e., the bulk motion of the gas is not sufficient to explain the observed line broadening. 

\begin{figure*}[!htb]
\centering
\includegraphics[angle=90,width=0.75\textwidth]{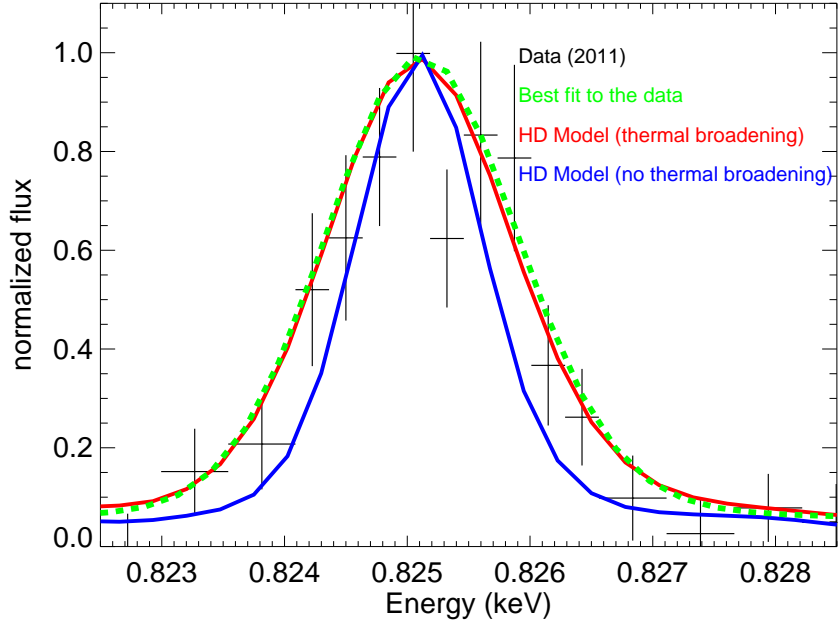}
\caption{Fe XVII line profile derived by our hydrodynamic model for 2011 with (red curve) and without (blue curve) thermal broadening, together with the corresponding 2011 $Chandra$ data (black crosses) and the Gaussian best fit to the data (green dashed curve).}
\label{fig:Feline}
\end{figure*}

\begin{figure*}[!htb]
\centering
\includegraphics[angle=90,width=0.75\textwidth]{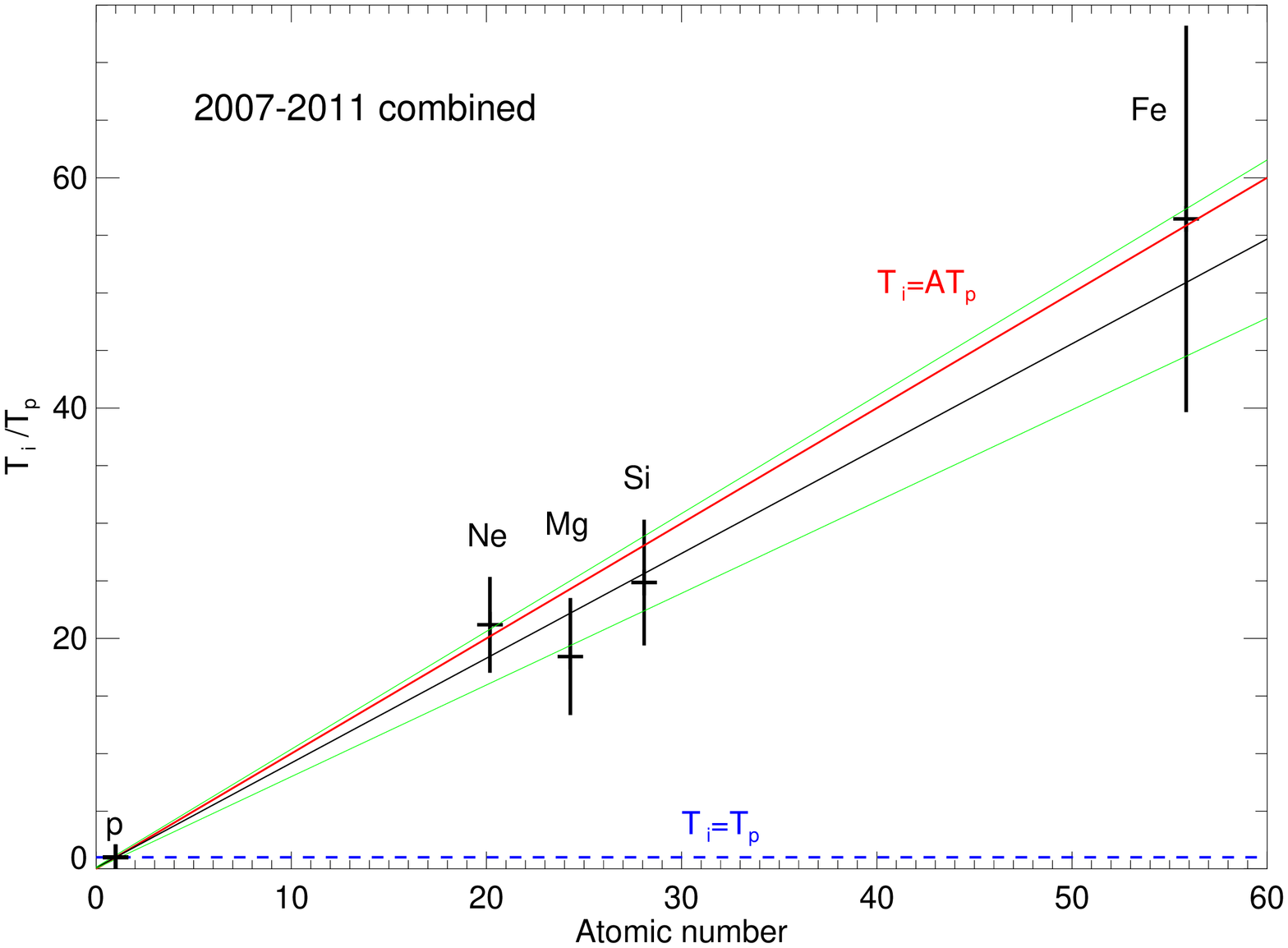}
\caption{Ion to proton temperature ratios measured by combining the 2007 and 2011 observations of SN 1987A for Ne, Mg, Si, and Fe lines (with corresponding $90\%$ error bars). The red line marks the mass-proportional trend predicted by equation (\ref{eq:tshock}), while the black line shows the best fit trend, with the corresponding $90\%$ confidence range indicated by green lines.}
\label{fig:results}
\end{figure*}

The line widths derived from the model by including also the thermal broadening are indicated by the red crosses in Fig. \ref{fig:broad} and are in excellent agreement with those observed, for all the ions, and for the 2007 and 2011 observations (see also Fig. \ref{fig:Feline}). 

To quantitatively check whether equation (\ref{eq:tshock}) holds for the different ions, we derived the post-shock ion temperatures $T_{i}(Ne,Mg,Si,Fe)$ through the differences between the observed broadening and the Doppler broadening of the corresponding emission lines (black and blue crosses in Fig. \ref{fig:broad}). The corresponding proton temperatures can be derived from our model. We found that the ion to proton temperature ratio increases monotonically with the ion mass both for the 2007 and the 2011 data sets. Given that the two observations provide consistent results, we combined them, by also combining the results from different ionization states of the same elements to further reduce the error bars.

Figure \ref{fig:results} shows the post-shock Ne, Mg, Si, and Fe temperatures normalized to the corresponding proton temperatures. We performed a simple linear regression on the data points in the figure, finding that the ion to proton temperature ratio $T_i/T_p$ increases with the ion mass $A$ as $T_i/T_p=kA$ with $k=0.90\pm0.12$. Therefore, the ion post-shock temperature is consistent with being mass-proportional. This result is in agreement with predictions of hybrid simulations of collisionless shocks\cite{cys17}.

In summary, through the combination of high-resolution X-ray spectra and 3-D hydrodynamic modeling, we ascertained the physical origin of the observed line profiles in SN 1987A, by pinpointing the roles of Doppler and thermal broadening. Our results unequivocally show that post-shock temperatures increase linearly with particle mass over a wide range of masses, previously unexplored. This is a validation of equation (\ref{eq:tshock}) for ions and probes the ion heating mechanism of collisionless shocks.
By analyzing multi-epoch observations performed at different phases of the shock-ring interaction, we also showed that the mass-proportional heating mechanism holds for different shock parameters.

\bigskip
\begin{addendum}
\item The software used in this work was, in part, developed by the U.S. Department of Energy-supported Advanced Simulation and Computing/Alliance Center for Astrophysical Thermonuclear Flashes at the University of Chicago. We acknowledge that the results of this research have been achieved using the PRACE Research Infrastructure resource MareNostrum III based in Spain at the Barcelona Supercomputing Center (PRACE Award N.2012060993). The scientific results reported in this article are based to a significant degree on data obtained from the Chandra Data Archive. SO, MM, GP, FB acknowledge financial contribution from the agreement ASI-INAF n.2017-14-H.O.

  \item[Corresponding author] Correspondence to M. Miceli~(email: marco.miceli@unipa.it).

\end{addendum}

\end{document}